\documentstyle[graphicx,aps,amstex,epsf]{revtex}
\begin{document} 
\twocolumn 
\draft \preprint{ }
\title{Kinetic Theory of Collective Excitations and Damping in Bose-Einstein
Condensed Gases} \author{U.  Al Khawaja and H.  T.  C.  Stoof} \address{Institute
for Theoretical Physics, University of Utrecht, Princetonplein 5,3584 CC
Utrecht, The Netherlands} 
\date{\today} 
\maketitle 
\begin{abstract} 
We calculate the frequencies and damping rates of the low-lying collective modes 
of a Bose-Einstein condensed gas at nonzero temperature. 
We use a complex nonlinear Schr\"odinger equation to determine the dynamics of 
 the condensate atoms, 
and couple it to a Boltzmann equation for the 
noncondensate atoms. In this manner we take into account both collisions between  
noncondensate-noncondensate and 
condensate-noncondensate atoms. 
We solve the linear response of these equations, 
using a time-dependent gaussian trial function for the condensate wave function 
and a truncated power expansion for the deviation function of the thermal cloud. 
As a result, our calculation turns out to be  characterized by two 
dimensionless parameters proportional 
to the noncondensate-noncondensate and condensate-noncondensate mean collision times. 
We find in general quite good agreement with experiment, both for the frequencies and 
damping of the collective modes. 
\end{abstract}

\pacs{PACS numbers: 03.75.Fi, 03.65.Db, 05.30.Jp, 32.80.Pj}

\section{Introduction}

The nonzero-temperature dependence of the frequencies and damping of collective modes
in trapped atomic Bose gases has been investigated extensively both 
experimentally \cite{{jila1},{jila2},{ket},{ket1}} and theoretically 
\cite{{notes},{griffin.stringari},{kagan},{griffin.hutchinson},%
{keith},{hutch},{giorgini},{griffin.low},{dodd},{griffin.zgn},%
{ho},{george.hydro},{george.coll},%
{griffin.zgnp},{griffin.new},{george.long},{stoof.m},{stoof.low},{odelin},{usama}}.  
Above the Bose-Einstein transition
temperature the lowest-lying collective modes have been calculated in the
collisionless regime using a collisionless Boltzmann or Landau-Vlasov 
equation \cite{george.coll}, 
and in the hydrodynamic regime either by using the conservation laws of hydrodynamics 
\cite{notes} or by using the appropriate quantum kinetic equation 
\cite{griffin.stringari}. 
In the collisionless regime the frequency of the collective mode is large compared
to the mean collision frequency, in contrast to the situation in the
hydrodynamic regime.  Several papers have also studied the regime
intermediate between the collisionless and hydrodynamic regimes by 
taking into account
interatomic collisions \cite{{george.coll},{odelin},{usama}}, 
which turn out to mostly lead to mode damping. These papers have
indicated that the experiments were performed under conditions intermediate
between collisionless and hydrodynamic. (This is particularly true for the experiments 
performed by Mewes{ \it et al.} \cite{ket} and Stamper-Kurn {\it et al.} \cite{ket1}). 
For temperatures far below the transition temperature 
the collisionless modes can be described accurately by the 
time-dependent Gross-Pitaevskii equation \cite{pitaevskii}. 
At higher temperatures the noncondensate fraction becomes substantial and the 
modes of the condensate  are now coupled to those of the 
thermal cloud. 
In Refs. \cite{{griffin.hutchinson},{keith},{giorgini}} 
the temperature dependence of the mode frequencies has been calculated 
by employing the Popov approximation to include 
the static mean-field effects of the noncondensate atoms 
in the Gross-Pitaevskii equation. 
This was improved later by taking into account also the dynamics of the thermal 
cloud using a RPA approximation 
\cite{hutch} or the collisionless Boltzmann equation \cite{stoof.m}. 
For the hydrodynamic regime a two-fluid model has been developed in 
Refs. \cite{{griffin.zgn},{ho}}. Moreover, the theory of Zaremba, Griffin, and 
Nikuni \cite{griffin.zgn} was 
improved later by the same authors to include collisions between condensate and 
noncondensate atoms \cite{griffin.zgnp}.

Using a combination of two previous papers, namely Refs. \cite{stoof.m} 
and \cite{usama}, 
we aim in this paper at 
interpolating between the collisionless and hydrodynamic regimes for 
experiments below the transition 
temperature. The authors of Ref. \cite{usama} 
have already performed such an interpolation for a gas above the critical temperature 
by using a  Boltzmann equation with a relaxation 
time approximation. 
In contrast, the authors of Ref. \cite{stoof.m} 
use a collisionless Boltzmann equation for the thermal cloud 
coupled to a time-dependent nonlinear Schr\"odinger equation for the condensate 
to consider the collisionless dynamics below the critical temperature. 
Here, we thus combine these two approaches by 
using again the Boltzmann equation and time-dependent 
nonlinear Schr\"odinger equation, but now including also 
the effects of interatomic collisions in the manner as put forward in 
Refs. \cite{griffin.zgn} and \cite{stoof.low}. 
This implies that we have to add two collision terms to the Boltzmann equation. 
The first collision term represents collisions between two noncondensate 
atoms and the second 
describes collisions between a condensate and a noncondensate atom. We use for 
both these collision terms a relaxation time approximation, since this 
approximation leads above the transition temperature 
to a good agreement with microscopic calculations as well 
as with experimental data  \cite{usama}. 
Furthermore, for consistency reasons we also have to include a damping term in the 
time-dependent non-linear Schr\"odinger equation, which is 
due to collisions between the condensate 
and noncondensate atoms. 
As a result our calculation will essentially be characterized by two parameters, namely 
$\tau_{22}$ and $\tau_{12}$, denoting the mean collision time for collisions between the 
noncondensate atoms and between condensate and noncondensate atoms, respectively. 
Note that we use here the same notation as in Ref. \cite{griffin.zgnp}. 
This will allow us to investigate the collisionless and hydrodynamic limits with 
respect to both $\tau_{22}$ and $\tau_{12}$,   
and enable us to fit 
our results for frequencies and damping with the experimental data.

To solve the complicated nonlinear dynamics of the gas, 
we employ a gaussian trial function for the wave function of the condensate atoms 
with three complex time-dependent variational 
parameters.   
For the thermal cloud we use a distribution function that incorporates 
deviations from the 
Bose-Einstein distribution function. The deviation function is a 
truncated power expansion 
in the momenta and coordinates of the atoms.  
Our system of coupled equations is then assembled from the Euler-Lagrange equations  
for the variational parameters in the condensate wavefunction 
and from all the linear and quadratic 
moments of the Boltzmann equation 
for the distribution function of the noncondensate atoms. 
The solution of the linearized version of this system of coupled equations 
results in 
a dispersion relation that gives the frequencies and damping of the coupled modes 
in terms of $\tau_{22}$ and $\tau_{12}$. Although we restrict our 
calculation to the axially 
symmetric traps used in the experiments, the generalization to 
fully anisotropic 
traps is straightforward. 

The rest of this paper is organized as follows: In the next section 
we present the theoretical details of our calculation. Specifically, 
we write down the time-dependent nonlinear Schr\"odinger equation for 
the condensate 
wave function and the linearized Boltzmann equation for the thermal cloud, 
and show how to treat the two collision 
terms in a relaxation time approximation. 
In section \ref{trial} we present our trial functions for the condensate wave 
function and noncondensate distribution function 
and in section \ref{ground} we discuss how we obtain the equilibrium state of the gas 
around which we have to expand to find the collective modes. 
In section \ref{frequencies} we present the resulting dispersion relation and 
discuss its collisionless and 
hydrodynamic limits. We then compare our results 
with experiment. 
We end in section \ref{conclusion} by summing up our main conclusions.

\section{Coupled Dynamics of the condensate and the thermal cloud}

It has been shown that the coupled time-dependent nonlinear Schr\"odinger equation 
and the quantum Boltzmann equation can both be derived starting 
from the equation of motion 
\cite{{giorgini},{griffin.zgnp},{stoof.low}}
\begin{eqnarray}
i\hbar{\partial\hat\psi({\bf r},t)\over {\partial t}}
&=&\text{\Huge{(}}
-{\hbar^2\nabla^2\over{2m}}
+V^{ext}({\bf r}) \nonumber\\
&+&g\hat\psi^\dagger({\bf r},t)\hat\psi({\bf r},t)
\text{\Huge{)}}
\hat\psi({\bf r},t)
\label{schrodinger},
\end{eqnarray}
for the Heisenberg field operator $\hat\psi({\bf r},t)$. Here, 
$g$ is the effective two-body interaction, which is given in 
terms of the scattering length $a$ and the atomic mass $m$ as $g=4\pi a\hbar^2/m$. 
The external potential we take here is a harmonic potential of the general form,
\begin{equation}
V^{ext}({\bf r})={1\over2}m(\omega_1^2x^2+\omega_2^2y^2+\omega_3^2z^2)
\label{V},
\end{equation}
where $\omega_i$ is the characteristic frequency of the trap in the $i$th direction. 
To obtain the time-dependent nonlinear Schr\"odinger equation we write 
$\hat{\psi}({\bf r},t)$ as
\begin{equation}
\hat\psi({\bf r},t)=\Phi({\bf r},t)+\hat{\psi}^\prime({\bf r},t),
\end{equation} 
where $\Phi({\bf r},t)$ is the appropriate nonequilibrium expectation value of 
$\hat\psi({\bf r},t)$ and the operator $\hat{\psi}^\prime({\bf r},t)$ describes the 
noncondensate atoms. The number density of condensate atoms $n_c({\bf r},t)$ 
is related to $\Phi({\bf r},t)$ by $n_c({\bf r},t)=|\Phi({\bf r},t)|^2$ 
while the number density of the noncondensate atoms $n^\prime$ equals 
$\left<{\hat{\psi^\prime}}^\dagger({\bf r},t){\hat{\psi^\prime}}
({\bf r},t)\right>$, where 
the brackets represent 
the averaging over the nonequilibrium density matrix. 
Substituting this form for $\hat{\psi}({\bf r},t)$ in the equation of motion 
and averaging, we obtain
\begin{eqnarray}
i\hbar{\partial\Phi({\bf r},t)\over {\partial t}}
&=&\text{\Huge{(}}
-{\hbar^2\nabla^2\over{2m}}
+V^{ext}({\bf r})
+gn_c({\bf r},t)
+2g{n^\prime}({\bf r},t)\nonumber\\
&-&iR({\bf r},t)
{\over}\text{\Huge{)}}
\Phi({\bf r},t),
\label{schrodinger+damping}
\end{eqnarray}
where $R({\bf r},t)$ is given by
\begin{eqnarray}
R({\bf r},t)
&=&-g\left(
\left<{\hat{\psi^\prime}}({\bf r},t){\hat{\psi^\prime}}({\bf r},t)
\right>\Phi^*({\bf r},t)\right.\nonumber\\
&+&\left.\left<{\hat{\psi^\prime}}^\dag({\bf r},t){\hat{\psi^\prime}}
({\bf r},t){\hat{\psi^\prime}}({\bf r},t)\right>
\right)
/\Phi({\bf r},t)
\label{r}.
\end{eqnarray}

Next, the quantum Boltzmann equation can be derived by writing an equation of motion 
for the distribution function of the noncondensate atoms $f({\bf p},{\bf r},t)$. 
This is usually done by writing $f({\bf p},{\bf r},t)$ as a Wigner transform
of $\left<{\hat{\psi^\prime}}^\dag({{\bf r}^\prime},t){\hat{\psi^\prime}}
({\bf r},t)\right>$ 
and then determining the 
time evolution of $f({\bf p},{\bf r},t)$ from the equation of motion 
in Eq. (\ref{schrodinger})
\cite{{griffin.zgnp},{stoof.low},{dorfman}}. 
In the Hartree-Fock approximation the resulting 
Boltzmann equation takes the form
\begin{equation}
\left[
\frac{\partial f}{\partial t}
+{\bf \nabla}_{\bf p}E\cdot{\bf \nabla}_{\bf r}
-{\bf \nabla}_{\bf r}E\cdot{\bf \nabla}_{\bf p}
\right]f
=C_{22}[f]+C_{12}[f]
\label{boltzmann},
\end{equation}
where $C_{22}$ is the contribution to the rate of change of $f$ 
due to collisions between noncondensate atoms, 
 while $C_{12}$ is the contribution due to collisions 
between the condensate and the noncondensate atoms. 
 The energy $E({\bf p},{\bf r},t)$ of the noncondensate atoms 
 is in this approximation given by \cite{stoof.m}
\begin{equation}
E({\bf p},{\bf r},t)
=p^2/2m+V^{ext}({\bf r})+2gn({\bf r},t)
\label{e.thermal},
\end{equation}
where $n=n_c+n^\prime$ is the total density. 
 
In Ref. \cite{griffin.zgnp} explicit forms of the two collision terms 
have been written down as follows
\begin{eqnarray}
C_{22}[f]
&=&{2g^2\over (2\pi)^5\hbar^7}
\int d{\bf p}_2\int d{\bf p}_3\int d{\bf p}_4
\delta({\bf p}+{\bf p}_2-{\bf p}_3-{\bf p}_4)\nonumber\\
&\times&\delta(E+E_2-E_3-
E_4)\nonumber\\
&\times&\left[
(1+f)(1+f_2)f_3f_4-f f_2(1+f_3)(1+f_4)
\right]
\label{c22},
\end{eqnarray}   
and
\begin{eqnarray}
C_{12}[f]
&=&{2g^2n_c\over (2\pi)^2\hbar^4}
\int d{\bf p}_1\int d{\bf p}_2\int d{\bf p}_3
\nonumber\\
&\times&\delta(m{\bf v}_c+{\bf p}_1-{\bf p}_2-{\bf p}_3)\nonumber\\
&\times&\delta(E_c+E_1-E_2-
E_3)\nonumber\\
&\times&\left[
\delta({\bf p}-{\bf p}_1)-\delta({\bf p}-{\bf p}_2)-\delta({\bf p}-{\bf p}_3)
\right]\nonumber\\
&\times&
\left[
(1+f_1)f_2f_3-f_1(1+f_2)(1+f_3)
\right]
\label{c12}.
\end{eqnarray}   
Here $E_c({\bf r},t)$ and ${\bf v}_c({\bf r},t)$ are the local energy and velocity of 
the condensate atoms, 
 $E_i$ is a shorthand for $E({\bf p}_i,{\bf r},t)$,  
and similarly $f_i$ is a shorthand for $f({\bf p}_i,{\bf r},t)$. 
As is shown in Ref. \cite{griffin.zgnp}, the conservation of the total number of atoms 
constraints the collision term $C_{12}$ in the Boltzmann equation 
to be related to the damping term $R({\bf r},t)$ in the complex nonlinear 
Schr\"odinger equation as
\begin{equation}
2n_c({\bf r},t)R({\bf r},t)=
\int{d{\bf p}\over(2\pi)^3}C_{12}[f]
\label{rc1}.
\end{equation}

The coupled equations given in Eqs. (\ref{schrodinger+damping}), (\ref{boltzmann}), 
and (\ref{rc1}) in principle fully determine the dynamics of the Bose-Einstein condensed 
gas in the Hartree-Fock approximation. 
Since the dynamics of the thermal cloud is experimentally only important for 
temperatures 
which are larger or comparable to the mean-field interactions 
we believe that this is sufficiently accurate and we do not need to use the Popov 
approximation at first instance. 
Nevertheless, these equations are too complicated to be solved 
exactly and some approximation 
is called for.

\subsection{Trial Functions}
\label{trial}

For the noncondensate atoms we start 
by linearizing the Boltzmann equation in small 
deviations of the distribution function around its equilibrium value, namely
\begin{equation}
f({\bf p},{\bf r},t)=f^{(0)}({\bf p},{\bf r})
+f^{(0)}({\bf p},{\bf r})(1+f^{(0)}({\bf p},{\bf r}))\psi({\bf p},{\bf r},t)
\label{psi},
\end{equation}
where $f^{(0)}({\bf p},{\bf r})$ is the equilibrium distribution function given by
\begin{equation}
f^{(0)}({\bf p},{\bf r})
=\left\{
\exp{\left[(E({\bf p},{\bf r})-\mu)/k_BT\right]}
-1\right\}^{-1}
\label{f0}.
\end{equation}
Here $\mu$ is the chemical potential and 
$k_B$ is Boltzmann's constant. 
As a result the linearized Boltzmann equation takes the form
\begin{equation}
\left[
\frac{\partial}{\partial t}
+{\bf \nabla}_{\bf p}E\cdot{\bf \nabla}_{\bf r}
-{\bf \nabla}_{\bf r}E\cdot{\bf \nabla}_{\bf p}
\right]\psi
=C_{22}[\psi]+C_{12}[\psi]
\label{boltzmann.linearized}.
\end{equation}

It is shown in Ref. \cite{usama} that for an uncondensed gas 
a trial function for $\psi({\bf p},{\bf r},t)$ 
of the form 
\begin{eqnarray}
\psi &=&A_{1}x^{2}+B_{1}xp_{x}+C_{1}p_{x}^{2}+A_{2}y^{2}+B_{2}yp_{y}
\nonumber\\
&+&C_{2}p_{y}^{2}
+A_{3}z^{2}+B_{3}zp_{z}+C_{3}p_{z}^{2}
\label{psi.trial},
\end{eqnarray}
where $A_i(t)$, $B_i(t)$, and $C_i(t)$ are nine time-dependent functions, 
is appropriate to describe the low-lying collisionless breathing modes  
with frequencies $2\omega_i$, some of which have been observed experimentally. 
We will therefore assume the above expansion to be also reasonably accurate 
below the critical temperature.   
For the collision integrals $C_{22}$ and $C_{12}$ we use a relaxation 
time approximation. In such an approximation one associates these collision integrals 
to mean relaxation times. However, the approximate expressions for 
$C_{22}$ and $C_{12}$ should still take into account the conservation laws associated 
with the collision processes exactly. In the case of $C_{22}$, which represents collisions 
between the noncondensate atoms, the number of atoms in the thermal cloud, 
their total momentum and their total kinetic energy 
should all be conserved. 
In our ansatz for $\psi$, 
terms like $x^2$ and $xp_x$ should therefore not be affected by such collisions since they 
correspond to two collision invariants, namely the number of atoms and the total momentum 
in the $x$ direction, 
respectively. 
On the other hand, terms like $p_x^2$ will be affected by collisions since it is only 
the sum $p^2=p_x^2+p_y^2+p_z^2$ which is conserved during the collision process. 
Therefore, we can write the following expression for the linear operator 
$C_{22}$ \cite{{george.long},{usama}} 
\begin{eqnarray}
C_{22}[\psi_i]&=&-{1\over\tau_{22}}
\left\{
\begin{array}{ccc}
(\psi_i-p^2/3)&,&\;\;\;\;\psi_i=p_x^2, p_y^2, p_z^2\\
0 &,&\;\;\;\;\;\rm{otherwise}
\end{array}
\right.,\nonumber\\&&
\label{c22.relaxation}
\end{eqnarray}
where $\tau_{22}$ is a mean collision time for the 
noncondensate-noncondensate atomic collisions. 
Note that in this expression $C_{22}[p^2]=0$, 
ensuring the conservation of the total kinetic energy. 

For $C_{12}$, which represents collisions between the condensate 
and the noncondensate atoms, the number of atoms is not conserved, since the 
collision process involves transport of atoms 
back and forth from the condensate into the thermal cloud. 
This statement means mathematically that the zeroth moment of $C_{12}$, i.e., 
$\int d{\bf p}C_{12}$, 
does not vanish in 
contrast to the case of $C_{22}$ where $\int d{\bf p}C_{22}=0$. 
As a first attempt to associate with $C_{12}$ a mean collision time $\tau_{12}$ 
for noncondensate-condensate atomic collisions, we may write 
$C_{12}\propto1/\tau_{12}$. However, we observe from 
Eq. (\ref{c12}) that $C_{12}\propto n_c({\bf r})$. 
This dependence on $n_c({\bf r})$ requires us to assign to $\tau_{12}$ 
a position dependence that follows from $n_{c}({\bf r})$ as 
\begin{equation}
{1\over\tau_{12}({\bf r})}={1\over\tau_{12}({\bf 0})}{n_c({\bf r})\over n_c({\bf 0})}
\label{tau12.position},
\end{equation} 
where $\tau_{12}({\bf 0})$ and $n_c({\bf 0})$ are the noncondensate-condensate 
mean atomic collision time and condensate density at the center 
of the trap, respectively. 
Generally, $\tau_{12}({\bf 0})$ may, just like $\tau_{22}$, 
also have temperature dependence. 
By multiplying and dividing $1/\tau_{12}({\bf r})$ by $\hbar g$ we can thus 
take $C_{12}$ as obeying
\begin{equation}
C_{12}\propto \alpha {2gn_c({\bf r})\over\hbar}
\label{c12.alfa},
\end{equation}
where we introduced the dimensionless constant 
\begin{equation}
\alpha={\hbar\over 2gn_{c}({\bf 0})\tau({\bf 0})}
\label{alfa}.
\end{equation}
This form for $C_{12}$ is convenient for a reason that becomes clear 
lateron in this section. 
It should be emphasized here that the dimensionless 
parameter $\alpha$ is the second free parameter in our phenomenological 
calculation. The first dimensionless parameter being $1/{\bar{\omega}
}\tau_{22}$ where $\bar{\omega}=\sqrt[3]{\omega_1\omega_2\omega_3}$. 
Therefore, the collisionless and hydrodynamic regimes, as well as the  
intermediate regime, will be described by only these two parameters.   
Finally, we make the relation between 
$C_{12}$ and $\alpha$ precise by taking 
\begin{eqnarray}
C_{12}[\psi_i]&=&-\alpha{2gn_c({\bf r})\over\hbar}\nonumber\\
&\times&\left\{
\begin{array}{ccc}
(\psi_i-p^2/3)&,&\;\psi_i={p_x}^2, {p_y}^2, {p_z}^2\\
\psi_i&,&\;\psi_i={x}^2, {y}^2, {z}^2\\
0 &,&\;\rm{otherwise}
\end{array}
\right.
\label{c12.relaxation}.
\end{eqnarray}
This completes our description of the treatment of the thermal cloud. 
Next we have to consider the condensate.

It is known that a time-dependent gaussian 
ansatz for the condensate wave function gives the correct 
frequencies of the lowest modes at zero temperature 
\cite{{carcia},{castin},{stoof.stat}}. 
Furthermore, the gaussian ansatz has also been used in Ref. \cite{stoof.m} 
for the whole temperature range below the transition temperature 
and leads to rather good agreement with experiment. 
Therefore, we employ here again a gaussian ansatz for the 
wave function of the condensate. It has the following form
\begin{eqnarray}
\Phi({\bf r},t)
&=&\sqrt[4]{8N_cb_{1r}b_{2r}b_{3r}\over\pi^{3}}
\exp{\left[-(b_1x^2+b_2y^2+b_3z^2)\right]},\nonumber\\&&
\label{phi}
\end{eqnarray}  
where $b_1$, $b_2$, and $b_3$ are complex time-dependent variational parameters and 
$b_{1r}$, $b_{2r}$, and $b_{3r}$ are their real parts, respectively. 
Similarly, we denote the imaginary parts, 
which will appear lateron, as $b_{1i}$, $b_{2i}$, and $b_{3i}$. 
The prefactor of $\Phi({\bf r},t)$ guarantees its normalization 
$\int d{\bf r}|\Phi({\bf r},t)|^2$ to 
be equal to the number of condensate 
atoms $N_c$. 

To obtain the first set of our coupled equations of motion, 
we start by writing down the energy functional that corresponds 
to the nonlinear Schr\"odinger equation in Eq. (\ref{schrodinger+damping}), namely
\begin{eqnarray}
E_c[\Phi]&=&\int d{\bf r}
\text{\Huge{[}}
{\hbar^2\left|{\bf \nabla \Phi}\right|^2\over 2m}
+V^{ext}n_c
+{1\over2}g({n_c}^2+4n_cn^\prime)\nonumber\\ 
&-& iRn_c
 {\over}\text{\Huge{]}}
\label{energyfunctional}
\end{eqnarray}
with $n_c=|\Phi|^2$. 
We evaluate this energy functional using the expression for $\Phi$ from 
Eq. (\ref{phi}), and by noting that the density of the thermal cloud
\begin{equation}
n^\prime=\int {d{\bf p}\over(2\pi)^3}f({\bf p},{\bf r},t)
\label{ntilde},
\end{equation}
and the damping term $R({\bf r},t)$ can be calculated 
using Eqs. (\ref{rc1}) and (\ref{c12.relaxation}). 
We notice here that $-iR({\bf r},t)n_c$ turns out to be proportional 
to the Hartree-Fock interaction 
term in Eq. (\ref{energyfunctional}), with a proportionality constant 
equals to $-i\alpha$. This explains the reason for writing $C_{12}$ as 
in Eq. (\ref{c12.alfa}). 
The equations of motion for the condensate dynamics are now the 
Euler-Lagrange equations resulting from varying the 
lagrangian 
\begin{equation}
L={1\over2}i\hbar\int d{\bf r}
\left(
\Phi^*{\partial\over \partial t}\Phi
-\Phi{\partial\over \partial t}\Phi^*
\right)
-E_c[\Phi]
\label{lagrangian}
\end{equation}
with respect to the 6 variational parameters $b_{kr}$ and $b_{ki}$.

The second set involves the equations of motion for the constants $A_i$, $B_i$, 
and $C_i$. It is obtained from taking appropriate moments of the Boltzmann 
equation in Eq. (\ref{boltzmann.linearized}). 
In detail, the moments are calculated by multiplying this 
equation by $f^{(0)}({\bf p},{\bf r})$ 
and the various 
components of $\psi$ in Eq. (\ref{psi.trial}), and then 
integrating over $\bf p$ and $\bf r$. This results in 9 equations of motion. 
In combination with the previous 6 ones, 
we thus have 15 coupled equations of motion. The coupling is provided on the one hand 
by the mean-field interaction $2gn_cn^\prime$ and the imaginary damping 
term $iRn_c$ in the condensate energy functional 
$E_c$ in Eq. (\ref{energyfunctional}), 
and on the other hand by the contribution $2gn_c$ to the 
Hartree-Fock energy $E$ in Eq. (\ref{e.thermal}) 
and the $C_{12}$ collision term in the Boltzmann equation.

\subsection{Equilibrium}
\label{ground}

Finally, it is important to note that linearization of the equations 
of motion of the condensate is obligatory to be consistent 
with the equations of motion of the noncondensate part which are already 
linearized. 
Therefore, we need to calculate the equilibrium state of the gas. 
In principle, to obtain equilibrium properties we should minimize the free 
energy $F=E_{tot}-TS$, where $E_{tot}$ is the total energy and 
$S$ is the entropy of the gas, with respect to some variational 
parameters that characterize the 
widths of the condensate and the thermal cloud. 
However, a simplified estimate of the total energy and 
the entropy contribution to the free energy above the transition temperature 
shows that for the experimental conditions of interest the former 
is dominant over the later. 
This simplified estimate can be made by using a one-parameter gaussian ansatz for 
the thermal cloud density, namely $n^\prime\propto\exp(-x^2/R^2)$, 
where $R$ is the radius of the cloud. 
It turns out that in the Thomas-Fermi limit the entropy contribution to the free energy  
is of order $k_BT\ln\gamma^{3/5}$, where $\gamma=N^\prime a/\bar a$, 
$N^\prime$ is the total number of atoms in the thermal cloud,  
and ${\bar a}=\sqrt{\hbar/m\bar\omega}$ 
is the harmonic oscillator length,  
while the total energy contribution 
is of order $k_BT\gamma^{2/5}$. 
Therefore, if $\gamma\gg1$ the entropy contribution is much less than the 
total energy contribution. 

As we shall see in the next section, the experiments were performed with a 
temperature-dependent total number of atoms ranging from about $6000$ atoms 
at zero temperature to approximately $40000$ atoms at the transition temperature. 
If we extend this simple estimate to temperatures below the transition temperature 
it turns out that the condition $\gamma\gg1$ is satisfied only for high temperatures 
but not for low temperatures since the number of thermal atoms becomes small. 
However, this condition is then no longer important since the 
free energy of the thermal 
cloud at such temperatures is small compared to the condensate energy.
Therefore, we will in the following only minimize the total 
energy $E_{tot}$ with respect 
to the variational parameters of the condensate and the thermal cloud. 
To be able to calculate $E_{tot}$
we assume that the 
distribution function of the noncondensate atoms has the same form of that of 
a noninteracting gas but with varying spatial widths that effectively take  
into account the mean-field effects of both the noncondensate and condensate 
atoms. This effective distribution function is written as
\begin{eqnarray}
f^{(0)}_{eff}
&=&\left[
\exp
{
\left(
{p^2/2m-\mu\over k_BT}+{x^2\over{R_1}^2}+{y^2\over{R_2}^2}+{z^2\over{R_3}^2}
\right)
}
-1\right]^{-1},\nonumber\\&&
\label{f.effective}
\end{eqnarray}
where $R_1$, $R_2$, and $R_3$ are the widths of the noncondensate cloud in the 
three directions. 
The total energy is now a function of 6 variational parameters:
\begin{equation}
E_{tot}=E_{tot}(b_{1r}^{(0)},b_{2r}^{(0)},b_{3r}^{(0)},
R_1,R_2,R_3)
\label{e.effective},
\end{equation} 
where $b_{1r}^{(0)}$, $b_{2r}^{(0)}$, and $b_{3r}^{(0)}$ are the equilibrium values 
of $b_{1r}$, $b_{2r}$, and $b_{3r}$, respectively. 
The equilibrium is obtained by minimizing this energy with respect to these  
variational parameters. 

The results of such a minimization  will be shown in the next section,  
where we present the dispersion relation that results from solving 
the above-described system of 15 linearized coupled equations of motion. 
We discuss also the collisionless and hydrodynamic limits of these results.

\section{The Dispersion Relation: Frequencies and Damping Rates}
\label{frequencies}

Our calculation accounts, in fully anisotropic traps, for 9 modes of the gas. 
In axially symmetric traps this number reduces to 6 modes. 
These are the in-phase and out-of-phase combinations of the 
two monopole ($m=0$) modes  and one quadrupole ($m=2$) 
mode of both the condensate and the 
thermal cloud. 
Here we denote with $m$ the projection of the angular momentum 
of the mode along the axis of symmetry of the trap \cite{jila2}.   
We focus in this paper on the two lowest-lying $m=0$ and $m=2$ modes 
observed experimentally. 
It turns out that for the experimentally relevant temperature range, the 
in-phase $m=0$, and $m=2$ modes correspond mostly to oscillations of the 
thermal cloud, whereas the out-of-phase modes are mostly condensate oscillations. 
Therefore, we shall often refer in this paper to the in-phase modes as the 
thermal cloud or noncondensate modes and the out-of-phase modes as 
the condensate modes. 
Although our calculation provides results for another, higher-lying, monopole mode, 
we shall not discuss it further here. 
Moreover, we emphasize that throughout the following we perform our calculations 
for parameters taken from  
the experiments of Jin {\it et al.} \cite{jila2}, i.e.,  
with $^{87}$Rb atoms in an axially-symmetric trap with anisotropy ratio 
$\omega_3/\omega_1=\omega_3/\omega_2=\sqrt{8}$. 
 
\begin{figure}[p]
\begin{center}
\includegraphics[width=0.5\textwidth]{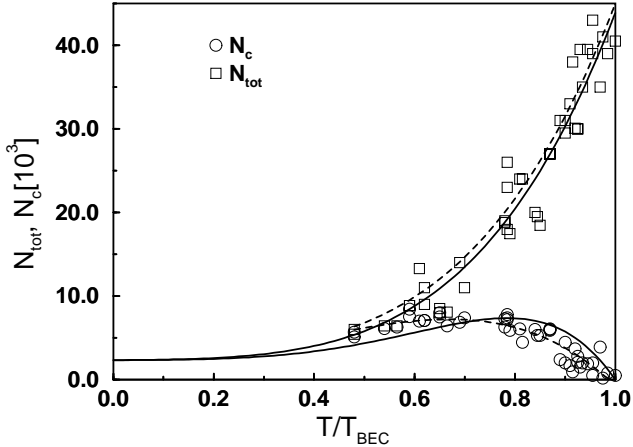}
\end{center}
\caption{
The measured total number of atoms $N_{tot}$ (squares) and the number of 
condensate atoms $N_c$ (circles). 
The dashed lines are a polynomial fit in powers of 
$T/T_{BEC}$ to the experimental data . 
The solid lines represent a fit to $N_{tot}$ only and using for 
$N_{c}$ the relation $N_c=N_{tot}(1-(T/T_{BEC})^2)$. 
The solid line fit of $N_{tot}$ is shifted slightly downwards 
in order to get better agreement with the data for $N_c$. 
The measured points were taken from Ref. [2].
}
\label{n}
\end{figure}

\begin{figure}[p]
\begin{center}
\includegraphics[width=0.5\textwidth]{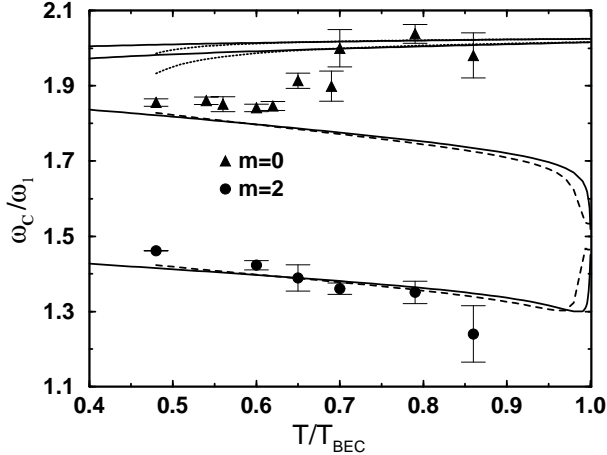}
\end{center}
\caption
{
Collisionless frequencies for $\alpha=0$, i.e., with 
no collisions between condensate and noncondensate 
atoms. The solid lines correspond to the approximate fit to 
the total and condensate number 
of atoms as shown by the solid lines in Fig. \ref{n}. 
The dashed lines correspond to our best fit in 
the temperature interval as is given by the dashed lines in Fig. \ref{n}. 
The points are the experimental measurements for the $m=0$ and $m=2$ condensate modes 
[2]. The labels of the theoretical curves in terms of the values of $m$ are shown in 
Fig. \ref{freqs2}.
}
\label{freqs1}
\end{figure}

In that particular experiment the measurements 
were performed with a temperature-dependent total number of 
atoms as a result of the loss of atoms during evaporative cooling. 
The total number of atoms $N_{tot}$, 
as well as the number of condensate atoms $N_c$, are measured 
in the temperature range $T/T_{BEC}\simeq0.48$ to $T/T_{BEC}\simeq1.0$, 
where $T_{BEC}$ is the Bose-Einstein transition temperature. 
These measurements can be easily fitted with polynomials in 
$T/T_{BEC}$ as shown by the dashed line in Fig. \ref{n}. 
An extrapolation of such a fit to temperatures below $0.48T_{BEC}$ 
leads, however, to nonphysical situations. 
In particular, the two curves for $N_{tot}$ 
and $N_c$ cross at least once before reaching zero temperature. 
We overcome this problem by fitting only $N_{tot}$ with the experimental data, and 
then using  $N_{c}$=$N_{tot}(1-(T/T_{BEC})^2)$ in analogy with the ideal gas relation 
$N_c$=$N_{tot}(1-(T/T_{BEC})^3)$ \cite{notes}. Fig. \ref{n} shows 
with the solid line the results of this slightly less accurate fit.  
It should be noted that using the ideal gas relation 
will grossly overestimate all the experimental points of $N_c$ 
for $T/T_{BEC}>0.7$. 
It turns out 
that the difference between the calculated frequencies 
using the less accurate fit from those calculated using the best fit are 
much smaller than the uncertainties in the measured frequencies 
for the experimental range of temperatures. 
This is shown in Fig. \ref{freqs1}. 
Since we want to show also results for the complete temperature interval 
from zero to $T_{BEC}$, we employ from now on always 
the former fit, i.e., the solid lines in Fig. \ref{n}. 

\begin{figure}[p]
\begin{center}
\includegraphics[width=0.5\textwidth]{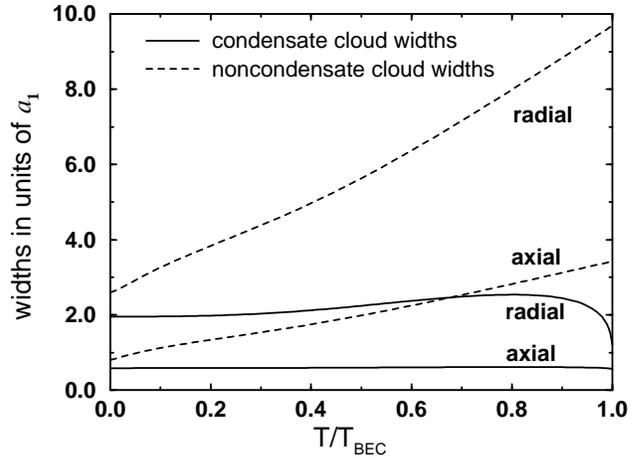}
\end{center}
\caption{
The equilibrium widths of the condensate and noncondensate clouds in the axially 
symmetric trap of Ref. [2]. 
In reference to the text, the radial width of the noncondensate cloud is $R_1 (=R_2)$, 
and the axial width is $R_3$. 
For the condensate cloud, the radial width is 
$(2b_{1r}^{(0)})^{-1/2} (=(2b_{2r}^{(0)})^{-1/2})$ and the axial width is 
$(2b_{3r}^{(0)})^{-1/2}$, as can be seen from Eq. (\ref{e.effective}). 
All lengths here are scaled to the harmonic oscillator 
 length $a_1=(\hbar/m\omega_1)^{1/2}$. 
}
\label{widths}
\end{figure}

Before starting with calculating 
the frequencies and damping rates of the collective modes, 
we show in Fig. \ref{widths} 
the result of the minimization of the total energy 
that is required to obtain the equilibrium conditions of the gas, as described in the 
previous section. 
We plot the equilibrium 
widths of both condensate and noncondensate clouds as a function of 
temperature. 
We notice from this figure that at zero temperature and at the transition temperature 
we obtain rather good estimates for the widths of the condensate and thermal clouds. 
At zero temperature, where the thermal cloud is absent, we observe that the radial 
width of the condensate equals approximately $2a_1$, a result that is consistent with 
the familiar property that 
due to the mean-field interactions the equilibrium condensate width 
is larger than the ideal gas result 
$a_1$. The axial width, which is roughly $1/\lambda^{1/2}$ of the radial width, 
is suppressed due to the anisotropy of the trap. 
We can also see in this figure that at the transition temperature 
the radial width of the condensate is slightly larger than the ideal gas one. 
This slight expansion is 
caused by the presence of the thermal cloud, whose mean-field interaction has the effect 
of slightly reducing the spring constants of the effective trapping potential.   
Returning to the problem of the collective mode frequencies and damping, 
we start by neglecting collisions between the condensate 
and noncondensate atoms, which means that $\alpha=0$. 
In this case the dispersion relation turns out to have the following general structure
\begin{equation}
iw
\left(
P_C^{(0)}(\omega\tau_{22})^2
+iP_I^{(0)}\omega\tau_{22}
-P_H^{(0)}
\right)=0
\label{dispersion.a=0},
\end{equation}        
where $P_C^{(0)}$, $P_I^{(0)}$, and $P_H^{(0)}$ are 6th order polynomials in $\omega^2$ 
which can all be factorized as
\begin{equation}
P_C^{(0)}
=\prod_{k=1}^{6}(\omega^2-\omega_{Ck}^2)
\label{pc01},
\end{equation}
\begin{equation}
P_I^{(0)}
=\prod_{k=1}^{6}(\omega^2-\omega_{Ik}^2)
\label{pc02},
\end{equation}
and
\begin{equation}
P_H^{(0)}
=\prod_{k=1}^{6}(\omega^2-\omega_{Hk}^2)
\label{pc03}.
\end{equation}
Here, $\omega_{Ck}$, $\omega_{Hk}$, and $\omega_{Ik}$, $k=1,2,3$, 
are temperature-dependent 
collisionless, intermediate, and hydrodynamic frequencies, respectively. 
The superscripts in $P_C^{(0)}$, $P_I^{(0)}$, and $P_H^{(0)}$ indicate the 
value of $\alpha$. 

The general structure of this dispersion relation agrees with the result of Refs. 
\cite{{odelin},{usama}} above the transition temperature, where this 
dispersion relation was studied in detail. 
Particularly, it was shown that damping rates calculated 
using this relation agree in order of 
magnitude with experiments and numerical calculations. 

\begin{figure}[p]
\begin{center}
\includegraphics[width=0.5\textwidth]{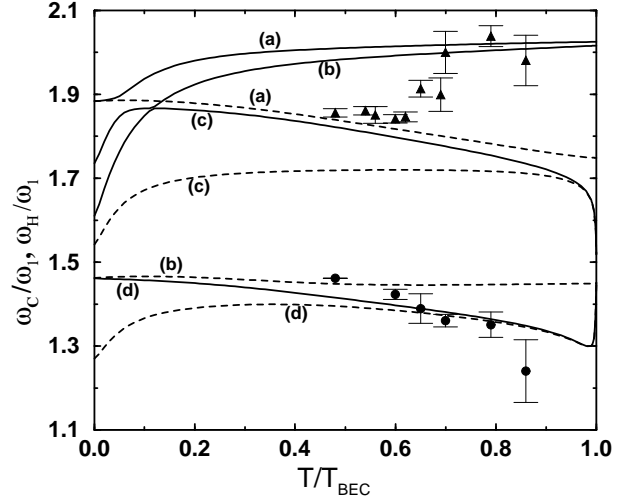}
\end{center}
\caption{
Collisionless (solid curves) and hydrodynamic (dashed curves) 
collective mode frequencies for 
$\alpha=0$. 
The curves are labeled in terms of the values of $m$ 
and the phase difference between the condensate and thermal oscillations 
$\theta$: 
(a) $m=0$, $\theta=0$, 
 (b) $m=2$, $\theta=0$,
 (c) $m=0$, $\theta=\pi$,
 (d) $m=2$, $\theta=\pi$.
}
\label{freqs2}
\end{figure}

The collisionless regime is defined by the condition $\omega\tau_{22}\gg1$. 
Therefore, we find from Eq. (\ref{dispersion.a=0}) that the collisionless frequencies 
are $\omega_{Ck}$. 
In the hydrodynamic limit $\omega\tau_{22}\ll1$, and the hydrodynamic 
frequencies are  $\omega_{Hk}$. 
Note that we use the word hydrodynamic here to denote that the thermal 
cloud is in the hydrodynamic 
regime though $\alpha=0$ and there are therefore 
no collisions between condensate and noncondensate atoms. 
In a sense this regime is thus precisely the limit discussed by Nikuni, 
Zaremba, and Griffin \cite{griffin.zgnp}.  
Using the experimental parameters we calculate these frequencies for the whole 
temperature range below $T_{BEC}$. 
In Fig. \ref{freqs2} we present the results of this calculation for the 
$m=0$ and $m=2$ modes together with the experimental data. 
In this figure the collisionless curves agree with those of
Bijlsma and Stoof \cite{stoof.m} for $T/T_{BEC}> 0.2$, which is 
not surprising since we use 
similar ansatz functions. The discrepancy for $T/T_{BEC}<0.2$ is due to the different 
ways of treating the equilibrium state. 

Above the transition temperature, the explanation of the mode damping
was based on the fact that the measured frequencies were less than the theoretical 
collsionless frequencies. 
This shift in frequency 
is then interpreted to be due to the fact that the system is shifted 
from the collisionless regime towards the hydrodynamic regime.  
Collisional damping associated with this shift was calculated and 
compared to the experimental damping. 
It is clear from Fig. \ref{freqs2} that below the transition temperature 
this kind of explanation is not possible, since most of the experimental points 
are not located between the collisionless and hydrodynamic curves. 
To investigate the possibility that this damping might be due to the 
condensate-noncondensate atomic collisions, we now study the effect of these collisions 
by taking $\alpha$ to be nonzero. 
For nonzero $\alpha$ the dispersion relation takes the general form
\begin{eqnarray}
\left(\sum_{k=0}^{9}P_{C}^{(k)}\alpha^k\right)(\omega\tau_{22})^2
&+&i\left(\sum_{k=0}^{8}P_{I}^{(k)}\alpha^k\right)\omega\tau_{22}
\nonumber\\&-&\sum_{k=0}^{7}P_{H}^{(k)}\alpha^k=0
\label{dispersion.a}.
\end{eqnarray}
We note that the $k=0$ term in this expression corresponds to
Eq. (\ref{dispersion.a=0}), where $\alpha$ was indeed taken to be zero. 
Frequencies and damping rates can now be obtained by 
writing $\omega$ in terms of a real and imaginary parts, 
$\omega=\omega_r+i\omega_i$, and then inserting this expression in the last equation. 
In general, one obtains expressions for $\omega_r$ and $\omega_i$ 
in terms of ${\bar{\omega}}\tau_{22}$ and $\alpha$. 

\begin{figure}[p]
\begin{center}
\includegraphics[width=0.5\textwidth]{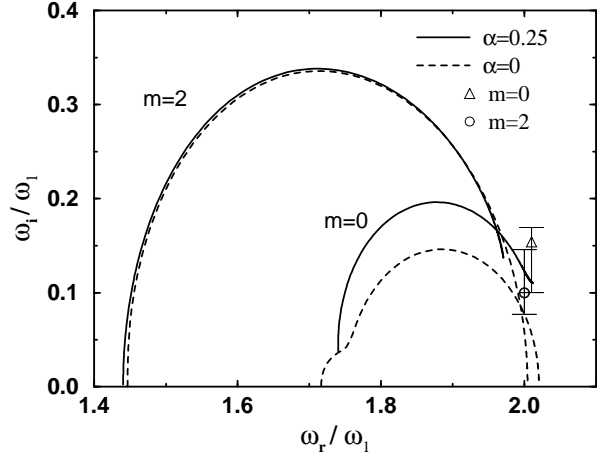}
\end{center}
\caption{
Damping rate $\omega_i$ versus frequency $\omega_r$ for noncondensate oscillations, 
at $T=.79T_{BEC}$ for $\alpha=0$ 
and $\alpha=0.25$. 
The two points correspond to the experimentally 
observed frequency and damping rate of the thermal 
cloud oscillation.  
}
\label{freqs3}
\end{figure}

We start our discussion of the 
frequencies and damping rates for $T/T_{BEC}=0.79$,  
since at this temperature the experiment provides data for both the condensate and 
noncondensate oscillations. 
We present the results of our calculation 
for the $m=0$ and $m=2$ modes of the thermal cloud  
in Fig. \ref{freqs3}, where we have eliminated ${\bar\omega}\tau_{22}$ 
and plotted $\omega_i$ as a function of $\omega_{r}$ for 
$\alpha=0$ and $\alpha=0.25$. 
On the same plot we show the two experimental points corresponding to the 
$m=0$ and $m=2$ modes of the noncondensate oscillation. 
In this figure, the right end of the curves represent the collisionless regime  
(${\bar\omega}\tau_{22}\gg1$) and the left end is 
the hydrodynamic regime (${\bar\omega}\tau_{22}\ll1$). Therefore, 
the location where the dashed curves intersect with the horizontal axis can also 
be seen in Fig. \ref{freqs2} at $T=0.79T_{BEC}$.
For the experimental point corresponding to the monopole mode, 
the error bars do not 
intersect with the dashed curve ($\alpha=0$), 
but they do with the solid curve ($\alpha=0.25$) 
indicating 
that a nonzero value of $\alpha$ is needed to account for the experimental damping. 
This indicates that condensate-noncondensate atomic collisions are mainly responsible  
for the observed damping of this mode. 
This is less clear for the quadrupole mode 
although the experimental data is certainly not inconsistent with this conclusion. 

\begin{figure}[p]
\begin{center}
\includegraphics[width=0.5\textwidth]{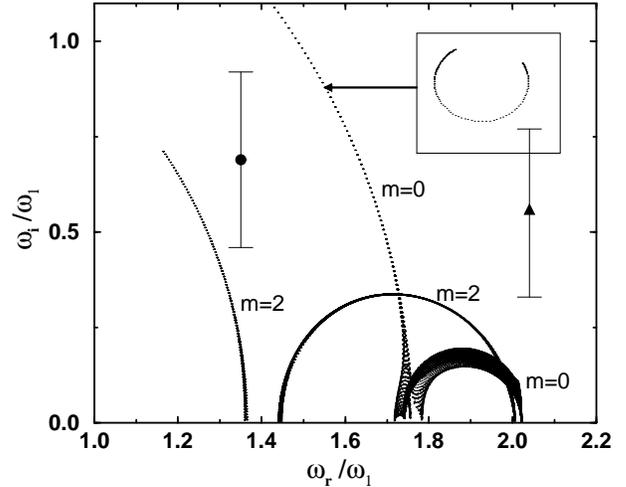}
\end{center}
\caption{
Damping rate $\omega_i$ versus frequency $\omega_r$ of condensate and 
noncondensate oscillations 
at $T=.79T_{BEC}$, for $\alpha$ changing from 0 to 0.25 in steps of 0.025. 
The two points correspond to the experimental measurements of the condensate oscillation.  
}
\label{freqs4}
\end{figure}

\begin{figure}[p]
\begin{center}
\includegraphics[width=0.4\textwidth]{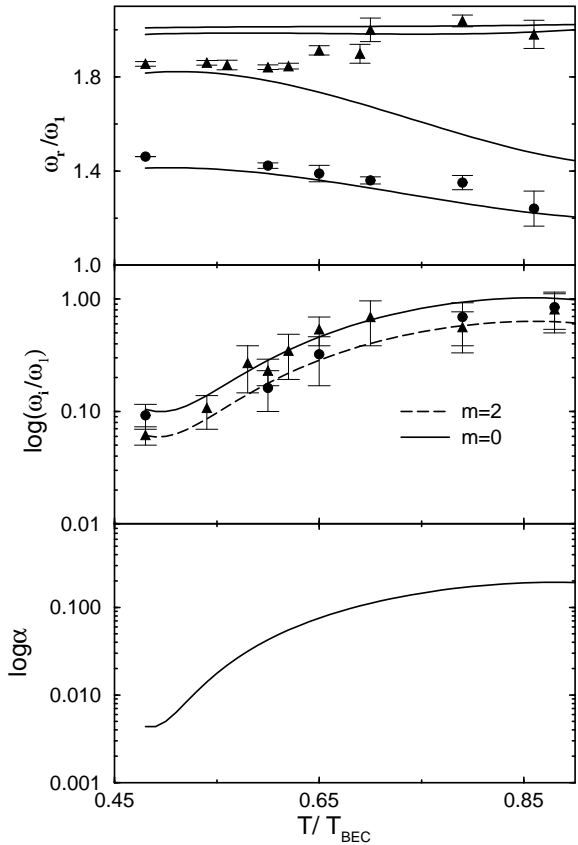}
\end{center}
\caption{
Fitting all the experimental data on the mode frequencies 
and damping rates with a 
single temperature-dependent function $\alpha$. 
}
\label{freqs6}
\end{figure}

\begin{figure}[p]
\begin{center}
\includegraphics[width=.45 \textwidth]{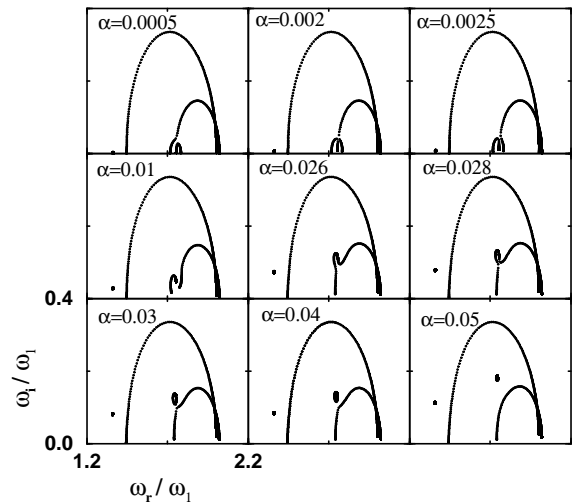}
\end{center}
\caption
{
Damping rate $\omega_i$ versus frequency $\omega_r$ of condensate and 
noncondensate oscillations 
at $T=.79T_{BEC}$ with $\alpha$ changing from 0.0005 to 0.05. 
The four curves in each plot correspond to the in-phase and out-of-phase 
$m=0$ and $m=2$ modes which are also shown in Fig. \ref{freqs4}. 
The present figure corresponds to the region in Fig. \ref{freqs4} 
where the in-phase and out-of-phase $m=0$ modes 
come close to each other. 
One can see here that the anticrossing feature of these two $m=0$ modes is not 
only present at $\alpha=0$ [21] but also 
for nonzero values of $\alpha$. 
}
\label{freqs5}
\end{figure}

Next we show in Fig. \ref{freqs4} at the same temperature 
both the condensate as well as the noncondensate $m=0$ and $m=2$ modes for values 
of $\alpha$ ranging from 0 to 0.25. 
The effect of $\alpha$ on the noncondensate modes is a slight shift and rotation of the 
curves but their general structure is roughly preserved. This can be seen more clearly in 
Fig. \ref{freqs3}. 
The effect of $\alpha$ on the condensate modes is much larger, 
it gives rise to a large upward shift for the whole curve. 
The effect of ${\bar{\omega}}\tau_{22}$ on these curves becomes smaller 
for larger values of $\alpha$, which can be seen by noting that the radius 
of the small semicircles becomes 
smaller for larger values of $\alpha$.
Again, we see that a nonzero value of $\alpha$ provides sufficient damping 
to account for the experimental observations. 
However, there is a discrepancy between our calculation and the experiment in 
the value of the frequency of the $m=0$ mode, 
since our calculation predicts $\omega/\omega_1\approx1.68\pm0.4$, whereas the 
experimental data give a frequency $\approx2.05\omega_1$. 
This discrepancy was also present in the calculations of Refs. \cite{{hutch},{stoof.m}} 
and is discussed in Ref. \cite{stoof.m}. 
The authors of the last reference suggested that a possible reason for this discrepancy may be 
that the observed mode is in fact an in-phase $m=0$ mode 
(the upper solid curve in Fig. \ref{freqs2}) 
rather than an out-of-phase one. 
This suggestion was based on calculating the oscillator strenghts of the in-phase  
and out-of-phase $m=0$ modes. It turned out that it is indeed possible experimentally to 
excite both modes simultaneously in the temperature 
range $T\approx0.25T_{BEC}$ to $T\approx0.5T_{BEC}$. 
Although the calculation in Ref. \cite{stoof.m} was performed in the collisionless limit, 
where there is  
no damping, we expect that with damping this argument remains qualitatively correct. 
For the rest of the experimental points we can, 
in principle, explain the data using a two-parameter fit, namely 
with $\alpha$ and ${\bar\omega}\tau_{22}$. 
In fact, the collisions between atoms from the condensate 
with those from the thermal cloud are the main cause of damping in the 
condensate modes, and   
we can even explain the data using a one-parameter fit, namely only $\alpha$, 
by assuming that the system is in the collisionless regime with respect to 
${\bar\omega}\tau_{22}$, i.e., ${\bar\omega}\tau_{22}\gg1$. 
We perform this calculation by using a function $\alpha(T/T_{BEC})$ that 
gives the best fit for the experimental damping rates. 
In Fig. \ref{freqs6} we plot the function $\alpha(T/T_{BEC})$ and the resulting 
temperature-dependent frequencies and damping rates. 
This figure contains the main results of this paper. First of all we notice that 
now, with a nonzero $\alpha$, we obtain better agreement with the experimental 
data for the quadrupole mode frequencies 
than before where $\alpha$ was taken to be zero. 
This can be seen by comparing this figure with Fig. \ref{freqs2}. 
We note also that, depending on the value of $\alpha$, the mode frequency 
may shift upwards or downwards. By comparing Fig. \ref{freqs2} with 
Fig. \ref{freqs6} for the mode frequencies 
one can clearly see that for temperatures below approximatelly 
$0.6T_{BEC}$ the theoretical curves are shifted upwards, whereas for the higher 
temperatures the curves are shifted downwards, in the end giving rise to the 
good agreement 
of the quadrupole mode with experiment.     
Secondly, we notice that the same function $\alpha(T/T_{BEC})$ also gives
a good fit for the damping rates of both the monopole and quadrupole modes. 
Finally, there is still a discrepancy between our predictions and the 
experimental findings 
for the frequencies of the monopole mode at higher temperatures, 
but this may be resolved by a reinterpretation of the observed 
mode, as explained previously.       
\\
\\
\\
Finally, we want to mention an interesting feature 
of the in-phase and out-of-phase $m=0$ modes which can already be seen in 
Fig. \ref{freqs2} at $T/T_{BEC}\approx0.05$. 
It is the familiar anticrossing tendency of these modes 
obtained also by Bijlsma and Stoof\cite{stoof.m} and 
Zaremba {\it et al.} \cite{griffin.zgnp}. It is more dramatic here than in these 
two works due to the presence of the new parameter $\alpha$. 
This behavior can be seen in Fig. \ref{freqs5} where we 
see the two modes come close to each other and 
eventually intersect at some `critical' 
values of $\alpha$ twice. 
\\
\\
\\
\\

\section{Conclusion}
\label{conclusion} 

We have extended the method of calculating collisional damping, used previously 
above the Bose-Einstein condensation transition temperature \cite{{odelin},{usama}}, 
to below the transition temperature. 
Furthermore, we have included the effect of noncondensate-condensate collisions.
By comparing with experiment, we conclude that it is presumably this collision process 
which is mainly responsible for the observed damping. 
Our theory provides a general dispersion relation that gives complex mode frequencies 
at a certain temperature as a function of two dimensionless parameters, namely  
$1/{\bar\omega}\tau_{22}$ and $\alpha$, that characterize the noncondensate-noncondensate 
and noncondensate-condensate atomic collisions, respectively. 
Our results for the fully collisionless frequencies 
agree with those of Ref. \cite{stoof.m} for most of the temperature 
range below the transition temperature. 
The $m=0$ and $m=2$ hydrodynamic 
frequencies are, to the best of our knowledge, calculated here  
below the transition temperature 
for the first time.

At present we have not carried out a microscopic calculation of the mean collision times 
$\tau_{22}$ and $\tau_{12}\propto\alpha^{-1}$, 
which are treated as phenomenological parameters here with the possibility   
to investigate the intermediate regime 
with respect to either of the two collision processes. 
However, the homogeneous calculations of Zaremba, Nikuni, and Griffin \cite{griffin.low} 
indicate that the function $\alpha(T/T_{BEC})$ that we have obtained from a fit 
to the experimental data, has the correct order of magnitude and qualitatively also 
the correct temperature-dependence. 
Nevertheless such a microscopic calculation for the inhomogeneous experimental conditions of 
interest would be very desirable and is left for future work.   

We obtain a rather good agreement with the experimental results for the mode frequencies 
and damping rates apart from the discrepancy 
for the $m=0$ out-of-phase mode 
for $T>0.7T_{BEC}$. While the experiment shows an upward shift in the frequency with 
respect to its zero temperature value, we predict a downward shift.  
Resolution of this discrepancy seems to require more accurate experimental data, 
which will also provide  
decisive comparison with the theory presented here.  
\section*{Acknowledgements}
The authors would like to thank Michiel Bijlsma, 
Chris Pethick and Henrik Smith for useful suggestions 
and remarks.

\end{document}